\documentclass[10pt,aps,prl,showpacs,twocolumn,amsmath,amssymb,superscriptaddress]{revtex4-1}
\usepackage{amssymb,amsthm,amsfonts,amstext}
\usepackage{color}
\usepackage{ifpdf}
\ifpdf 
  \usepackage[pdftex]{graphicx}
  \DeclareGraphicsExtensions{.pdf,.png,.jpg,.jpeg,.mps}
  \usepackage{pgf}
  \usepackage{tikz}
\else 
  \usepackage{graphicx}
  \DeclareGraphicsExtensions{.eps,.bmp}
  \usepackage{pgf}
  \usepackage{tikz}
  \usepackage{pstricks}
\fi

\DeclareMathOperator{\trace}{tr}

\begin{document}

\title{Quantum correlations require multipartite information principles}
\author{Rodrigo Gallego}
\affiliation{ICFO-Institut de Ci\`encies Fot\`oniques, E-08860 Castelldefels, Barcelona, Spain}
\author{Lars Erik W\"{u}rflinger}
\affiliation{ICFO-Institut de Ci\`encies Fot\`oniques, E-08860 Castelldefels, Barcelona, Spain}
\author{Antonio Ac\'in}
\affiliation{ICFO-Institut de Ci\`encies Fot\`oniques, E-08860 Castelldefels, Barcelona, Spain}
\affiliation{ICREA-Instituci\'o Catalana de Recerca i Estudis Avan\c{c}ats, Lluis Companys 23, 08010 Barcelona, Spain}
\author{Miguel Navascu\'es}
\affiliation{Department of Mathematics, University of Bristol, Bristol BS8 1TW, U.K.}

\begin{abstract}
Identifying which correlations among distant observers are
possible within our current description of Nature, based on
quantum mechanics, is a fundamental problem in Physics. Recently,
information concepts have been proposed as the key ingredient to
characterize the set of quantum correlations. Novel information
principles, such as information causality or non-trivial
communication complexity, have been introduced in this context and
successfully applied to some concrete scenarios. We show in this
work a fundamental limitation of this approach: no principle based
on bipartite information concepts is able to single out the set of
quantum correlations for an arbitrary number of parties. Our
results reflect the intricate structure of quantum correlations
and imply that new and intrinsically multipartite information
concepts are needed for their full understanding.
\end{abstract}

\maketitle

\paragraph*{Introduction}
An ubiquitous problem in Physics is to understand which
correlations can be observed among different events. In fact, any
theoretical model aims at predicting the experimental results of
measurements, or actions, performed at different space-time
locations. Naively, one could argue that any kind of correlations
are in principle possible within a general physical theory, and
that only the details of the devices used for establishing the
correlations imply limitations on them. Interestingly, this
intuition is not correct: general physical principles impose
non-trivial constraints on the allowed correlations among distant
observers, independently of any assumption on the internal working
of the devices. It is then a crucial question to identify
which correlations among distant observers are compatible with our
current description of Nature based on Quantum Physics. In
particular, it would be desirable to understand why some
correlations cannot be realized by quantum means, even if they do
not allow any faster-than-light communication~\cite{Popescu1994}.

Recently, information concepts have been advocated as the key
missing ingredient needed to single-out the set of quantum
correlations \cite{Dam2000,bub}. The main idea is to identify `natural' information
principles, formulated in terms only of correlations, which are
satisfied by quantum correlations and proven to be violated by
supra-quantum correlations. The existence of these supra-quantum
correlations, then, would have implausible consequences from an
information point of view. These information principles would
provide a natural explanation of why the correlations observed in
Nature have the quantum form. Celebrated examples of these
principles are information causality \cite{Pawlowski2009a} or
non-trivial communication complexity \cite{Dam2000,Dam2005}.
While the use of these information concepts has been
successfully applied to some specific scenarios
\cite{Brassard2006,Brunner2009,Allcock2009,Ahanj2010,Cavalcanti2010},
proving, or disproving, the validity of a principle for quantum
correlations is extremely challenging. On the one hand, it is
rather difficult to derive the Hilbert space structure needed for
quantum correlations from information quantities. On the other
hand, proving that some supra-quantum correlations are fully
compatible with an information principle seems out of reach, as
one needs to consider all possible protocols using these
correlations and show that none of them leads to a violation of
the principle. Thus, it is still open whether this approach is
able to fully determine the set of quantum correlations.

In this work, we consider a general scenario consisting of
an arbitrary number of observers and show a fundamental limitation
of this information-based program: no information principle based
on bipartite concepts is able to determine the set of quantum
correlations. Our results imply that determining the set of
quantum correlations for an arbitrary number of observers,
requires principles of an intrinsically multipartite structure.

\paragraph*{Non-signaling, local and quantum correlations}
The analyzed scenario consists of $n$ distant observers that can
perform $m$ possible measurements of $d$ possible results on their
systems. The observed correlations are described by the joint
probability distribution $P(a_1,\ldots,a_n|x_1,\ldots,x_n)$, where
$x_i=0,\ldots,m-1$ denotes the measurement performed by party
$i=1,\ldots,n$; and $a_i=0,\ldots,d-1$, the corresponding result.
Each system is just seen as a black box producing the output $a_i$
given the input $x_i$.

Consider first the situation in which the measurements by the
observers define space-like separated events. Then, the laws of
special relativity guarantee that no signal has been able to
propagate among the different observers. Under these conditions,
the statistics seen by a subset of  $k$ observers is independent
of any measurement performed by the other $n-k$ observers. Indeed,
if this were not the case, the $n-k$ observers could signal to the
remaining $k$ ones, even if they were causally disconnected.
Mathematically, the impossibility of faster-than-light
communication is imposed on the set of probabilities by requiring
that
\begin{multline}
\label{eq:nscorr}
    P(a_1\ldots a_k|x_1\ldots x_k)\\
=\sum_{a_{k+1}\ldots a_n}P(a_1\ldots a_n|x_1\ldots x_n)
\end{multline}
be independent of $x_{k+1},\ldots,x_n$. Similar relations hold for
any partition of the $n$ parties in two groups. These linear
constraints define the set of non-signaling correlations.

A subset of the non-signaling correlations is the set of
correlations having a local hidden variable model,
\begin{multline}
\label{clcorr} P_\mathrm{L}(a_1\ldots a_n|x_1\ldots
x_n)\\=\sum_\lambda p_\lambda P_1(a_1|x_1,\lambda)\ldots
P_n(a_n|x_n,\lambda) .
\end{multline}
These correlations are also called local or classical and
have a clear operational meaning: they can be established among
the observers when each of them produces locally the
outcome $a_i$ using the input $x_i$ and some pre-established
classical instructions, denoted by $\lambda$. As first shown by
Bell, they satisfy some non-trivial linear constraints, known as
Bell inequalities \cite{Bell1964}. It can also be shown
that some correlations are local if, and only if, they are
compatible with the no-signaling principle and
determinism~\cite{Valentini2002}. Indeed, they can always be
decomposed as mixtures of points where the result for each
measurement is assigned in a deterministic manner.

Quantum correlations correspond to those that can be obtained by
performing local measurements on an $n$-partite state. Formally,
one has
\begin{multline}
\label{qcorr} P_\mathrm{Q}(a_{1}\ldots a_{n}|x_{1}\ldots
x_{n})\\=\trace(\varrho\; M_{a_{1},x_1}^{(1)}\otimes \ldots
\otimes M_{a_{n},x_n}^{(n)}),
\end{multline}
where $\varrho$ is the $n$-partite quantum state and
$M_{a_i,x_i}^{(i)}$ the measurement operator by party $i$ yielding
outcome $a_i$ given measurement choice $x_i$. Quantum correlations
are known to lie between the set of classical and general
non-signaling correlations as there exist quantum correlations
which violate a Bell inequality and therefore have no classical
analog \cite{Bell1964}, and non-signaling correlations which are
supra-quantum \cite{Popescu1994},i.e., they cannot be written in
the form~\eqref{qcorr}. Despite having a clear mathematical
definition~\eqref{qcorr}, the set of quantum correlations lacks a
nice interpretation in terms of general principles, contrary to
the classical and non-signaling counterparts. As said, it has been
suggested that information concepts could provide the missing
principles for quantum correlations.

It is worth mentioning before proceeding with the proof of the
results that most of the existing examples of information
principles have been formulated in the bipartite scenario. For
example, information causality considers a scenario in which a
first party, Alice, has a string of $n_A$ bits. Alice is then
allowed to send $m$ classical bits to a second party, Bob.
Information causality bounds the information Bob can gain on the
$n_A$ bits held by Alice whichever protocol they implement making
use of the pre-established bipartite correlations and the message
of $m$ bits. Alice and Bob can violate this principle when they
have access to some supra-quantum correlations
\cite{Pawlowski2009a}. In the case $m=0$, information causality
implies that in absence of a message, pre-established correlations
do not allow Bob to gain any information about any of the bits
held by Alice, which is nothing but the no-signaling principle.
The multipartite version of the no-signaling principle consists in
the application of its bipartite version to all possible
partitions of the $n$ parties into two groups, see
\eqref{eq:nscorr}. This suggests the following generalization of
information causality to an arbitrary number of parties: given
some correlations $P(a_1,\ldots,a_n|x_1,\ldots,x_n)$, they are
said to be compatible with information causality whenever all
bipartite correlations constructed from them satisfy this
principle. This generalization ensures the correspondence between
no-signaling and information causality when $m=0$ for an arbitrary
number of parties. This generalization of information causality
has recently been applied to the study of extremal tripartite
non-signaling correlations~\cite{singapore}.

Regarding non trivial communication complexity, it studies how
much communication is needed between two distant parties to
compute probabilistically a function of some inputs in a
distributed manner. It can also be interpreted as a generalization
of the no-signaling principle, as it imposes constraints on
correlations when a finite amount of communication is allowed
between parties. Different multipartite generalizations of the
principle have been studied, see \cite{Buhrman1999}. However, as
for information causality, one can always consider the
straightforward generalization in which the principle is applied
to every partition of the $n$ parties in two groups.

\paragraph*{Supra-quantum correlations fulfilling information principles}

In this work, we show that any physical principle that, similarly
to no-signaling, is applied to every bipartition in the
multipartite scenario is not sufficient to characterize the set of
quantum correlations. We show this by finding tripartite
correlations that, on one hand, fulfill any information principle
based on bipartite concepts and, on the other hand, are
supra-quantum.

To grant that our distributions are compatible with any bipartite
information principle, we will restrict our search to a set of
tripartite correlations which behave classically under any system
bipartition. Let $P(a_1a_2a_3|x_1x_2x_3)$ be a non-signaling
tripartite distribution. We say that $P(a_1a_2a_3|x_1x_2x_3)$
admits a \emph{time-ordered bi-local (TOBL)} model if it can be
written as
\begin{equation}
\begin{aligned}
\label{eq:tobl}
  P&(a_1 a_2 a_3|x_1 x_2 x_3)\\
  &= \sum_\lambda p_\lambda^{i|jk} P(a_i|x_i, \lambda) P_{j\rightarrow k}(a_j a_k|x_j x_k,\lambda) \\
  &= \sum_\lambda p_\lambda^{i|jk} P(a_i|x_i, \lambda) P_{j\leftarrow k}(a_j a_k|x_j
  x_k,\lambda)
\end{aligned}
\end{equation}
for $(i,j,k)= (1,2,3), (2,3,1),(3,1,2)$, with the distributions
$P_{j\rightarrow k}$ and $P_{j\leftarrow k}$ obeying the
conditions
\begin{align}
 \label{eq:timeordered1}
  &P_{j \rightarrow k}(a_j|x_j,\lambda) = \sum_{a_k} P_{j \rightarrow k}(a_j a_k|x_j x_k,\lambda),\\
  \label{eq:timeordered2}
  &P_{j \leftarrow k}(a_k|x_k,\lambda) = \sum_{a_j} P_{j \leftarrow k}(a_j a_k|x_j x_k,\lambda).
\end{align}
The notion of TOBL correlations first appeared
in~\cite{Pironio2011} (see~\cite{tobl} for a proper introduction
and further motivation for such a models). As can be seen from
relations \eqref{eq:timeordered1} and \eqref{eq:timeordered2} we
impose the distributions $P_{j\rightarrow k}$ and $P_{j\leftarrow
k}$ to allow for signaling at most in one direction, indicated by
the arrow, see Table \ref{tab:point}. 

\begin{table}[ht]
 \centering
 \begin{center}
\begin{tabular}{c|c|c|c}
$x_2$ & $x_3$ & $a_2$ & $a_3$ \\
\hline
\hline
$0$ & $0$ & $0$ & $0$\\
\hline
$0$ & $1$ & $0$ & $1$\\
\hline
$1$ & $0$ & $1$ & $1$\\
\hline
$1$ & $1$ & $1$ & $0$\\
 \end{tabular}
\qquad
\begin{tabular}{c|c|c|c}
$x_2$ & $x_3$ & $a_2$ & $a_3$ \\
\hline
\hline
$0$ & $0$ & $0$ & $0$\\
\hline
$0$ & $1$ & $1$ & $1$\\
\hline
$1$ & $0$ & $0$ & $0$\\
\hline
$1$ & $1$ & $1$ & $1$\\
 \end{tabular}
\qquad
\begin{tabular}{c|c|c|c}
$x_2$ & $x_3$ & $a_2$ & $a_3$ \\
\hline
\hline
$0$ & $0$ & $1$ & $0$\\
\hline
$0$ & $1$ & $0$ & $1$\\
\hline
$1$ & $0$ & $0$ & $1$\\
\hline
$1$ & $1$ & $1$ & $0$\\
 \end{tabular}
 \end{center}
\caption{Different examples of deterministic bipartite probability
distributions $P_{23}(a_2 a_3|x_2x_3,\lambda)$ characterized by
output assignments to the four possible combination of
measurements. Left: inputs and outputs corresponding to a point
$P_{2 \rightarrow 3}(a_2 a_3|x_2x_3,\lambda)$ in the decomposition
\eqref{eq:tobl}. Center: inputs and outputs corresponding to a
point $P_{2 \leftarrow 3}(a_2 a_3|x_2x_3,\lambda)$ in
\eqref{eq:tobl}. Right: inputs and outputs corresponding to a
distribution which allows signaling in the two
directions.}
 \label{tab:point}
\end{table}

To understand the operational meaning of these models,
consider the bipartition $1|23$ for which systems $2$ and
$3$ act together. In this situation, $P(a_1a_2a_3|x_1x_2x_3)$ can
be simulated if a classical random variable $\lambda$
with probability distribution $p^{1|23}_\lambda$ is shared
by parts $1$ and the composite system $2-3$, and they
implement the following protocol: given $\lambda$, $1$ generates
its output according to the distribution $P(a_1|x_1, \lambda)$;
on the other side, and depending on which of the parties
$2$ and $3$ measures first, $2-3$ uses either $P_{2 \rightarrow
3}(a_2 a_3|x_2 x_3,\lambda)$ or $P_{2 \leftarrow 3}(a_2 a_3|x_2
x_3,\lambda)$ to produce the two measurement outcomes. Likewise,
any other bipartition of systems 1,2,3 admits a classical
simulation.

By construction, the set of tripartite TOBL models is convex and
includes all tripartite probability distributions of the form
(\ref{clcorr}). Moreover, it becomes classical under
postselection: indeed, suppose that we are given a tripartite
distribution $P(a_1a_2a_3|x_1x_2x_3)$ satisfying condition
(\ref{eq:tobl}), and a postselection is made on the outcome
$\tilde{a}_3$ of measurement $\tilde{x}_3$ by party $3$. Then one
has
\begin{eqnarray}
  &&\hspace{-10pt}P(a_1a_2|x_1x_2\tilde{x}_3\tilde{a}_3) \nonumber\\
  &&\hspace{10pt}=\sum_\lambda p'_\lambda P(a_1|x_1,\lambda)P'(a_2|x_2, \lambda),
  \label{post_sel}
\end{eqnarray}

\noindent with

\begin{eqnarray}
&&p'_\lambda=\frac{p_\lambda^{1|23}}{P(\tilde{a}_3|\tilde{x}_3)}P_{2\leftarrow 3}(\tilde{a}_3|\tilde{x}_3,\lambda), \nonumber\\
&&P'(a_2|x_2,\lambda)=P_{2\leftarrow 3}(a_2|x_2\tilde{x}_3\tilde{a}_3, \lambda).
\end{eqnarray}

\noindent Postselected tripartite TOBL boxes can thus be regarded
as elements of the TOBL set with trivial outcomes for one of the
parties.

\begin{figure}
\begin{picture}(0,0)%
\includegraphics{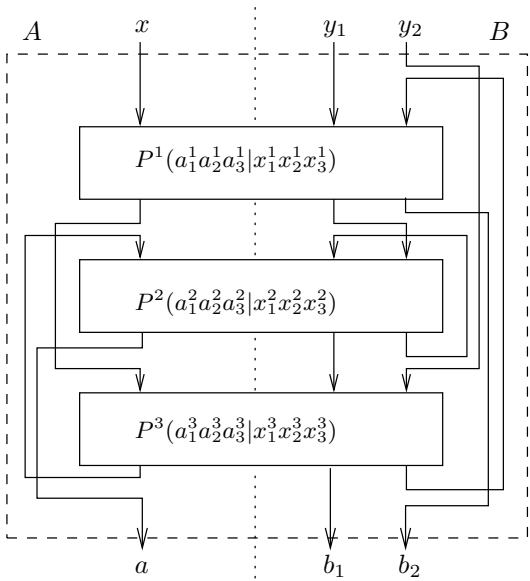}%
\end{picture}%
\setlength{\unitlength}{4144sp}%
\begingroup\makeatletter\ifx\SetFigFont\undefined%
\gdef\SetFigFont#1#2#3#4#5{%
  \reset@font\fontsize{#1}{#2pt}%
  \fontfamily{#3}\fontseries{#4}\fontshape{#5}%
  \selectfont}%
\fi\endgroup%
\begin{picture}(3138,3496)(4309,505)
\put(5086,2999){\makebox(0,0)[lb]{\smash{{\SetFigFont{9}{10.8}{\familydefault}{\mddefault}{\updefault}{\color[rgb]{0,0,0}$P^{1}(a_1^1a_2^1a_3^1|x_1^1x_2^1x_3^1)$}%
}}}}
\put(6661,3809){\makebox(0,0)[lb]{\smash{{\SetFigFont{10}{12.0}{\familydefault}{\mddefault}{\updefault}{\color[rgb]{0,0,0}$y_2$}%
}}}}
\put(5086,569){\makebox(0,0)[lb]{\smash{{\SetFigFont{10}{12.0}{\familydefault}{\mddefault}{\updefault}{\color[rgb]{0,0,0}$a$}%
}}}}
\put(5086,3809){\makebox(0,0)[lb]{\smash{{\SetFigFont{10}{12.0}{\familydefault}{\mddefault}{\updefault}{\color[rgb]{0,0,0}$x$ }%
}}}}
\put(6211,3809){\makebox(0,0)[lb]{\smash{{\SetFigFont{10}{12.0}{\familydefault}{\mddefault}{\updefault}{\color[rgb]{0,0,0}$y_1$}%
}}}}
\put(6211,569){\makebox(0,0)[lb]{\smash{{\SetFigFont{10}{12.0}{\familydefault}{\mddefault}{\updefault}{\color[rgb]{0,0,0}$b_1$}%
}}}}
\put(6661,569){\makebox(0,0)[lb]{\smash{{\SetFigFont{10}{12.0}{\familydefault}{\mddefault}{\updefault}{\color[rgb]{0,0,0}$b_2$}%
}}}}
\put(4411,3764){\makebox(0,0)[lb]{\smash{{\SetFigFont{10}{12.0}{\rmdefault}{\mddefault}{\updefault}{\color[rgb]{0,0,0}$A$}%
}}}}
\put(7201,3764){\makebox(0,0)[lb]{\smash{{\SetFigFont{10}{12.0}{\rmdefault}{\mddefault}{\updefault}{\color[rgb]{0,0,0}$B$}%
}}}}
\put(5086,1379){\makebox(0,0)[lb]{\smash{{\SetFigFont{9}{10.8}{\familydefault}{\mddefault}{\updefault}{\color[rgb]{0,0,0}$P^{3}(a_1^3a_2^3a_3^3|x_1^3x_2^3x_3^3)$}%
}}}}
\put(5086,2144){\makebox(0,0)[lb]{\smash{{\SetFigFont{9}{10.8}{\familydefault}{\mddefault}{\updefault}{\color[rgb]{0,0,0}$P^{2}(a_1^2a_2^2a_3^2|x_1^2x_2^2x_3^2)$}%
}}}}
\end{picture}%
\caption{\label{fig:wirings}Wiring of several tripartite
correlations distributed among parties $A$ and $B$. The generated
bipartite box accepts a bit $x$ (two bits $y_1,y_2$) as input on
subsystem $A$ ($B$) and returns a bit $a$ (two bits $b_1,b_2$) as
output. Relations \eqref{eq:tomodels} guarantee that the final
bipartite distribution $P_\mathrm{fin}(a,(b_1,b_2)|x,(y_1,y_2))$
admits a local model.}
\end{figure}

As mentioned in the introduction, to demonstrate that a set of
correlations is compatible with an information principle one needs
to consider all possible protocols using these correlations and
ensure that the correlations obtained this way are in accordance
with the principle. The most general protocol consists in
distributing an arbitrary number of boxes described by $P^1,
P^2,\ldots, P^N$ among three parties which are split into two
groups, $A$ and $B$. Both groups can process the classical
information provided by their share of the $N$ boxes. For
instance, outputs generated by some of the boxes can be used as
inputs for other boxes, see figure~\ref{fig:wirings}. This local
processing of classical information is usually referred to as
\textit{wirings}~\cite{wirings}. Thus, in order to prove
our result in full generality, we should consider all possible
wirings of tripartite boxes. We show next that if $P^1,
P^2,\ldots, P^N$ are in TOBL, then the resulting correlations
$P_{\mathrm{fin}}$ obtained after any wiring protocol have a local
decomposition with respect to the bipartition $A|B$, and therefore
fulfill any bipartite information principle.

For simplicity, we illustrate our procedure for the wiring shown
in figure~\ref{fig:wirings}, where boxes $P^{1}, P^{2} , P^{3}$
are distributed between two parties $A$ and $B$, and party $A$
only holds one subsystem of each box. The construction is
nevertheless general: it applies to any wiring and also covers
situations where for some TOBL boxes party $A$ holds two
subsystems instead of just one (or even the whole box).

From (\ref{eq:tobl}) we have
\begin{equation}
\label{eq:tomodels}
\begin{aligned}
  P^{i}&(a^{i}_1a^{i}_2a^{i}_3| x^{i}_1x^{i}_2x^{i}_3)\\ &= \sum_{\lambda^i} p_{\lambda^i}^{i} P_1^{i}(a^{i}_1|x_1^{i}, \lambda^i)P_{2 \rightarrow 3}^{i}(a_2^{i}a_3^{i}|x^{i}_2x^{i}_3, \lambda^i)\\
&= \sum_{\lambda^i} p_{\lambda^i}^{i} P_1^{i}(a^{i}_1|x^{i}_1,
\lambda^i)P_{2 \leftarrow 3}^{i}(a^{i}_2a_3^{i}|x_2^{i}x_3^{i}, \lambda^i),
 \end{aligned}
\end{equation}
\noindent for $i= 1,2, 3$. Consider the first box that receives an
input, in our case subsystem $2$ of $P^{1}$. The first outcome
$a_2^{1}$ can be generated by the probability distribution $P_{2
\rightarrow 3}^{1}(a_2^{1},a_3^{1}|x_2^{1},x_3^{1},\lambda^{1})$
encoded in the hidden variable $\lambda^1$ that models these first
correlations. This is possible because for this decomposition
$a_2^{1}$ is defined independently of $x_3^{1}$, the input in
subsystem $3$. Then, the next input $x_3^{2}$, which is equal to
$a_2^{1}$, generates the output $a_3^{2}$ according to the
probability distribution $P_{2 \leftarrow
3}^{2}(a_2^{2},a_3^{2}|x_2^{2},x_3^{2},\lambda^{2})$ encoded in
$\lambda^2$. The subsequent outcomes $a_2^{i}$ and $a_3^{i}$ are
generated in a similar way. The general idea is that outputs are
generated sequentially using the local models according to the
structure of the wiring on $2-3$. Finally, subsystem $1$ can
generate its outputs $a^{i}$ by using the probability distribution
$P_{1}^{i}(a_1^{i}|x^{i},\lambda^{i})$. This probability
distribution is independent of the order in which parties $2$ and
$3$ make their measurement choices for any of the boxes. Averaging
over all hidden variables one obtains $P_\mathrm{fin}$. This
construction provides the desired local model for the final
probability distribution.


To show the absence of a quantum realization for some elements of
the TOBL set of correlations, we use the Bell inequality known as
`Guess Your Neighbor's Input' \cite{Almeida2010}
\begin{equation}\label{gyni}
\begin{split}
\mathbf{B}(P_{\mathrm{Q}})=&P_{\mathrm{Q}}(000|000)+P_{\mathrm{Q}}(110|011)\\&+P_{\mathrm{Q}}(011|101)+P_{\mathrm{Q}}(101|110) \leq 1.\\
\end{split}
\end{equation}
The inequality is defined in a scenario consisting of three
parties, who can perform two measurements of two outcomes.
Interestingly, the bound is the same both for classical and
quantum correlations \cite{Almeida2010}. That is, correlations
violating this inequality are supra-quantum.

We have now presented all the necessary ingredients to prove our
main result. To demonstrate the existence of supra-quantum
correlations that are compatible with any bipartite information
principle we maximize the expression \eqref{gyni} over TOBL
correlations. This optimization defines a linear program that can
be efficiently solved. Formally we have
\begin{equation}\label{eq:linprog}
\begin{aligned}
\mathbf{B}_{\mathrm{max}} =  \; &{\text{maximize}}\; \mathbf{B}(P) \\
&\text{subject to}\\
& P(a_1 a_2 a_3|x_1 x_2 x_3)\in\mathrm{TOBL}.
\end{aligned}
\end{equation}
The maximization yields a value of
$\mathbf{B}_{\mathrm{max}}=\frac{7}{6}$, implying the existence of
supra-quantum correlations in TOBL. Details of this probability distribution  attaining the maximum of $7/6$ and its TOBL decomposition can be found in the Supplemental Material of this article \cite{supplement}.

\paragraph*{Conclusion}
To summarize, we have shown that there exist tripartite
non-signaling correlations that fulfill the principles of
information causality and non-trivial communication complexity
although they do not belong to the set of quantum correlations.
The presented reasoning also applies to every other principle
applied to the bipartitions of a multipartite system. This result
provides a helpful insight for the formulation of a future
principle aiming at distinguishing between quantum and
supra-quantum correlations. In contrast to the no-signaling
principle, such a forthcoming principle will need to be an
intrinsically multipartite concept. This suggests that future
research should be devoted to the development of information
concepts of genuinely multipartite character. More specifically,
one could investigate which multipartite generalizations of non
trivial communication complexity can be considered intrinsically
multipartite, and furthermore, how to generalize information
causality for the case of multipartite communication protocols.

\textit{Note added} After completion of this work, an extremal
point of the tripartite non-signaling polytope which is
supra-quantum and in TOBL was reported in~\cite{singapore}.

\paragraph*{Acknowledgements}
This work was supported by the ERC starting grant PERCENT, the
European EU FP7 Q-Essence and QCS projects, the Spanish
FIS2010-14830 and Consolider-Ingenio QOIT projects, and the Philip
Leverhulme Trust.


\begin{thebibliography}{20}
\bibitem{Popescu1994} S. Popescu and D. Rohrlich, Foundations of Physics, {\bf 24,} 379 (1994).
\bibitem{Dam2000} W. van Dam, \emph{Nonlocality \& Communication complexity,}Ph.D. thesis, University of Oxford (2000).
\bibitem{bub}
R. Clifton, J. Bub and H. Halvorson, Foundations of Physics 33, 1561-1591 (2003).
\bibitem{Pawlowski2009a} M. Pawlowski, \emph{et al}, Nature {\bf 461,} 1101 (2009).
\bibitem{Dam2005} W. van Dam, arXiv:quant-ph/0501159 (2005).
\bibitem{Brassard2006} G. Brassard \emph{et al}, \prl {\bf 96,} 250401 (2006).
\bibitem{Brunner2009} N. Brunner and P. Skrzypczyk, \prl {\bf 102,} 160403 (2009).
\bibitem{Allcock2009} J. Allcock \emph{et al}, \pra {\bf 80,} 040103 (2009).
\bibitem{Ahanj2010} A. Ahanj \emph{et al} \pra {\bf 81,} 032103 (2010).
\bibitem{Cavalcanti2010} D. Cavalcanti, A. Salles and V. Scarani, Nat. Comm. {\bf 1,} 136 (2010).
\bibitem{Bell1964} J. S. Bell, Physics {\bf 1,} 195 (1964).
\bibitem{Valentini2002} A. Valentini, Phys. Lett. A {\bf 297,} 273 (2002).
\bibitem{singapore} T.H. Yang et al, arXiv:1108.2293v1 [quant-ph] (2011).
\bibitem{Buhrman1999} H. Buhrman \emph{et al}, \pra {\bf 60,} 2737 (1999).
\bibitem{Pironio2011} S. Pironio, J. D. Bancal and V. Scarani, J. Phys. A: Math. Theor. {\bf 44} 065303 (2011).
\bibitem{tobl} J. Barrett \emph{et al}, in preparation; R. Gallego \emph{et
al}, in preparation.
\bibitem{wirings} Jonathan Allcock et al,\pra {\bf 80}, 062107 (2009).
\bibitem{supplement} See supplemental material
\bibitem{Almeida2010} M. L. Almeida \emph{et al}, \prl {\bf 104,} 230404 (2010).
\end{thebibliography}

\section{Appendix}

This appendix presents a tripartite non-signaling probability distribution that attains the maximum of $7/6$ for the `Guess Your Neighbor's Input' inequality, as well as its TOBL decomposition.
To simplify notation, let us switch from $(a_1a_2a_3)$ to $(abc)$; and from $(x_1x_2x_3)$, to $(xyz)$. Now, consider the no-signaling tripartite probability distribution $P(a,b,c|x,y,z)$ given by the probabilities shown in Table \ref{tab:ptot}.

\begin{table}[h]
 \begin{center}
\begin{tabular}{|c|cccccccc|}
\hline
 & 000 & 001 & 010 & 011 & 100 & 101 & 110 & 111\\
\hline
000 & $\frac{2}{3}$&0&0&0&0&0&0&$\frac{1}{3}$\\
001&$\frac{1}{3}$&$\frac{1}{3}$&0&0&0&0&$\frac{1}{6}$&$\frac{1}{6}$\\
010&$\frac{1}{3}$&0&$\frac{1}{3}$&0&0&$\frac{1}{6}$&0&$\frac{1}{6}$\\
011&$\frac{1}{6}$&$\frac{1}{6}$&$\frac{1}{6}$&$\frac{1}{6}$&0&$\frac{1}{6}$&$\frac{1}{6}$&0\\
100&$\frac{1}{3}$&0&0&$\frac{1}{6}$&$\frac{1}{3}$&0&0&$\frac{1}{6}$\\
101&$\frac{1}{6}$&$\frac{1}{6}$&0&$\frac{1}{6}$&$\frac{1}{6}$&$\frac{1}{6}$&$\frac{1}{6}$&0\\
110&$\frac{1}{6}$&0&$\frac{1}{6}$&$\frac{1}{6}$&$\frac{1}{6}$&$\frac{1}{6}$&$\frac{1}{6}$&0\\
111&0&$\frac{1}{6}$&$\frac{1}{6}$&$\frac{1}{6}$&$\frac{1}{6}$&$\frac{1}{6}$&$\frac{1}{6}$&0\\
\hline
\end{tabular}
\end{center}
\caption{Tripartite probability distribution $P(abc|xyz)$ attaining the maximum of $7/6$ for the `Guess Your Neighbor's Input' inequality, where the rows correspond to the inputs $xyz$ and the columns to the outputs $abc$.}
\label{tab:ptot}
\end{table}

The value of the `Guess Your Neighbor's Input' inequality for $P(a,b,c|x,y,z)$ equals
\begin{equation}
 \mathbf{B}(P)= \frac{2}{3} +\frac{1}{6} +\frac{1}{6} +\frac{1}{6}=\frac{7}{6}\not=1,
\end{equation}

\noindent and thus $P(a,b,c|x,y,z)$ cannot be approximated by any quantum system. Next we will prove that $P(a,b,c|x,y,z)$ belongs to the TOBL set of correlations, and so it is compatible with any bipartite information principle.

First, notice that $P(a,b,c|x,y,z)$ is invariant under permutations of the three parties. It is therefore enough to show that it admits a decomposition of the form (\ref{eq:tobl}) for the partition $A|BC$. Along this bipartition, probability distributions appearing in the decomposition \eqref{eq:tobl} are such that the outcome $a$ only depends on the measurement choice $x$ for every given $\lambda$; let $a_x$ denote this outcome for $x=0,1$. Conditions \eqref{eq:timeordered1} and  \eqref{eq:timeordered2} tell us that for every $\lambda$ the marginal $P_{B \rightarrow C}(b|y, \lambda)$ is independent of $z$, and the marginal $P_{B \leftarrow C}(c|z, \lambda)$ is independent of $y$. Thus, for $B \rightarrow C$ we have that $b$ depends on $y$ and $c$ depends on both $z$ and $y$. The possible outcomes will then be denoted $b_y,c_{yz}$. Similarly, for $B \leftarrow C$, the possible outcomes are $b_{yz},c_z$. Tables \ref{tab:detpoints_BC} and \ref{tab:detpoints_CB} contain the output assignments corresponding to deterministic probability distributions together with the weights $p_\lambda$ for $A|B \rightarrow C$ and $A|B \leftarrow C$, respectively. Note that, in agreement with \eqref{eq:tobl}, the outcome assignments for $A$ and the weights $p_{\lambda}$ are the same for both decompositions.


\begin{table}[h]
\begin{center}
\begin{tabular}{|c|c|cccccccc|}
\hline
$\lambda$&$p_\lambda$ &$a_0$& $a_1$&$b_{0}$&$b_{1}$&$c_{00}$&$c_{01}$&$c_{10}$&$c_{11}$\\
\hline
\hline
1&1/12 & 0&0&0&1&0&1&0&1\\
2&1/12 &0&0&0&0&0&1&0&1\\
3&1/12 &0&0&0&0&0&0&0&1\\
4&1/12 &0&0&0&1&0&0&0&1\\
5&1/12 &0&1&0&1&0&0&0&0\\
6&1/12 &0&1&0&0&0&1&0&0\\
7&1/12 &0&1&0&0&0&0&0&0\\
8&1/12 &0&1&0&1&0&1&0&0\\
9&1/6&1&0&1&1&1&1&1&0\\
10&1/6&1&1&1&0&1&0&1&1\\
\hline
\end{tabular}
\end{center}
\caption{TOBL decomposition into deterministic probabibility distributions characterized by outcome assignments for the bipartition $A|BC$ in the case $A|B \rightarrow C$. For every $\lambda$ the outcome $a$ only depends on $x$, and $b$ only depends on $y$.}
\label{tab:detpoints_BC}
\end{table}

\begin{table}[h]
\begin{center}
\begin{tabular}{|c|c|cccccccc|}
\hline
$\lambda$&$p_\lambda$ &$a_0$& $a_1$&$b_{00}$&$b_{01}$&$b_{10}$&$b_{11}$&$c_0$&$c_{1}$\\
\hline
\hline
1&1/12&0&0&0&0&0&1&0&0\\
2&1/12&0&0&0&0&0&1&0&1\\
3&1/12&0&0&0&0&1&1&0&0\\
4&1/12&0&0&0&0&1&1&0&1\\
5&1/12&0&1&0&0&0&0&0&0\\
6&1/12&0&1&0&0&0&0&0&1\\
7&1/12&0&1&0&0&1&0&0&0\\
8&1/12&0&1&0&0&1&0&0&1\\
9&1/6&1&0&1&1&1&0&1&1\\
10&1/6&1&1&1&1&0&1&1&0\\
\hline
\end{tabular}
\end{center}
\caption{TOBL decomposition into deterministic probability distributions characterized by outcome assignments for the bipartition $A|BC$ in the case $A|B \leftarrow C$. For every $\lambda$ the outcome $a$ only depends on $x$, and $c$ only depends on $z$.}
\label{tab:detpoints_CB}
\end{table}

\noindent It is trivial to see that both tables indeed reproduce $P(a,b,c|x,y,z)$, and hence such a distribution belongs to the TOBL set.

\end{document}